\documentclass[a4paper,12pt]{article}
\pdfoutput=1
\usepackage{epsfig}
\usepackage{amssymb}
\usepackage{amsfonts}
\usepackage{amsmath}
\usepackage{euscript}
\usepackage{verbatim}
\usepackage{latexsym}
\usepackage{graphicx}
\usepackage{caption}
\usepackage{float}
\usepackage{ytableau}
\usepackage{subcaption}
\usepackage{wrapfig}
	\usepackage[T1]{fontenc}
	\usepackage{tikz}
	\usetikzlibrary{decorations.pathmorphing}
\usepackage{tikz}
\usetikzlibrary{shapes.geometric, arrows,patterns,snakes}
\tikzstyle{ellip} = [ellipse, minimum width=3cm, minimum height=1cm,text centered, draw=black]

\newskip\humongous \humongous=0pt plus 1000pt minus 1000pt

\newif\ifdtup

\allowdisplaybreaks[1]

\catcode`\@=11

\@addtoreset{equation}{section}

\def\@normalsize{\@setsize\normalsize{15pt}\xiipt\@xiipt
\abovedisplayskip 14pt plus3pt minus3pt%
\belowdisplayskip \abovedisplayskip
\abovedisplayshortskip \z@ plus3pt%
\belowdisplayshortskip 7pt plus3.5pt minus0pt}

\def\small{\@setsize\small{13.6pt}\xipt\@xipt
\abovedisplayskip 13pt plus3pt minus3pt%
\belowdisplayskip \abovedisplayskip
\abovedisplayshortskip \z@ plus3pt%
\belowdisplayshortskip 7pt plus3.5pt minus0pt
\def\@listi{\parsep 4.5pt plus 2pt minus 1pt
     \itemsep \parsep
     \topsep 9pt plus 3pt minus 3pt}}

\relax

\catcode`@=12

\catcode`\@=11

\def\section{\@startsection{section}{1}{\z@}{3.5ex plus 1ex minus
   .2ex}{2.3ex plus .2ex}{\large\bf}}

\def\SymBoxes#1#2#3#4{\newdimen\un@t \un@t#3%
\raisebox{#1}{\rule{#2\un@t}{#4}\hskip-#2\un@t
\@tempdimb\un@t \advance\@tempdimb by-#4\@tempcntb#2\relax%
\@whilenum{\@tempcntb>0}\do{
\rule{#4}{\un@t}\hskip\@tempdimb \advance\@tempcntb by\m@ne}%
\hskip-#2\un@t \rule[\un@t]{#2\un@t}{#4}%
\rule[\un@t]{#4}{#4}\hskip-#4
\rule{#4}{\un@t}}\hskip-#4}                

\begin{document}

\newcommand{\beq}{\begin{equation}}
\newcommand{\eeq}{\end{equation}}
\newcommand{\bea}{\begin{eqnarray}}
\newcommand{\eea}{\end{eqnarray}}
\newcommand{\beas}{\begin{eqnarray*}}
\newcommand{\eeas}{\end{eqnarray*}}
\newcommand{\defi}{\stackrel{\rm def}{=}}
\newcommand{\non}{\nonumber}
\newcommand{\bquo}{\begin{quote}}
\newcommand{\enqu}{\end{quote}}
\renewcommand{\(}{\begin{equation}}
\renewcommand{\)}{\end{equation}}
\def \eqn#1#2{\begin{equation}#2\label{#1}\end{equation}}
\def\IZ{{\mathbb Z}}
\def\IR{{\mathbb R}}
\def\IC{{\mathbb C}}
\def\IQ{{\mathbb Q}}
\def\de{\partial}
\def\Tr{ \hbox{\rm Tr}}
\def\H{ \hbox{\rm H}}
\def\HE{ \hbox{$\rm H^{even}$}}
\def\HO{ \hbox{$\rm H^{odd}$}}
\def\K{ \hbox{\rm K}}
\def\Im{ \hbox{\rm Im}}
\def\Ker{ \hbox{\rm Ker}}
\def\const{\hbox {\rm const.}}
\def\o{\over}
\def\im{\hbox{\rm Im}}
\def\re{\hbox{\rm Re}}
\def\bra{\langle}\def\ket{\rangle}
\def\Arg{\hbox {\rm Arg}}
\def\Re{\hbox {\rm Re}}
\def\Im{\hbox {\rm Im}}
\def\exo{\hbox {\rm exp}}
\def\diag{\hbox{\rm diag}}
\def\longvert{{\rule[-2mm]{0.1mm}{7mm}}\,}
\def\a{\alpha}
\def\dag{{}^{\dagger}}
\def\tq{{\widetilde q}}
\def\p{{}^{\prime}}
\def\W{W}
\def\N{{\cal N}}
\def\hsp{,\hspace{.7cm}}

\def\br{\nonumber\\}
\def\IZ{{\mathbb Z}}
\def\IR{{\mathbb R}}
\def\IC{{\mathbb C}}
\def\IQ{{\mathbb Q}}
\def\IP{{\mathbb P}}
\def \eqn#1#2{\begin{equation}#2\label{#1}\end{equation}}

\newcommand{\sgm}[1]{\sigma_{#1}}
\newcommand{\idd}{\mathbf{1}}

\newcommand{\C}{\ensuremath{\mathbb C}}
\newcommand{\Z}{\ensuremath{\mathbb Z}}
\newcommand{\R}{\ensuremath{\mathbb R}}
\newcommand{\rp}{\ensuremath{\mathbb {RP}}}
\newcommand{\cp}{\ensuremath{\mathbb {CP}}}
\newcommand{\vac}{\ensuremath{|0\rangle}}
\newcommand{\vact}{\ensuremath{|00\rangle}                    }
\newcommand{\oc}{\ensuremath{\overline{c}}}
\begin{titlepage}
\begin{flushright}
CHEP XXXXX
\end{flushright}
\bigskip
\def\thefootnote{\fnsymbol{footnote}}

\begin{center}
{\Large
{\bf Towards a Finite-$N$ Hologram
}
}
\end{center}

\bigskip
\begin{center}
{\large  Chethan KRISHNAN$^a$\footnote{\texttt{chethan.krishnan@gmail.com}}, and K. V. Pavan KUMAR$^a$\footnote{\texttt{kumar.pavan56@gmail.com}} \vspace{0.15in} \\ }
\vspace{0.1in}

\end{center}

\renewcommand{\thefootnote}{\arabic{footnote}}

\begin{center}
$^a$ {Center for High Energy Physics,\\
Indian Institute of Science, Bangalore 560012, India}

\end{center}

\noindent
\begin{center} {\bf Abstract} \end{center}
We suggest that holographic tensor models related to SYK are viable candidates for exactly (ie., non-perturbatively in $N$) solvable holographic theories. The reason is that in these theories, the Hilbert space is a spinor representation, and the Hamiltonian (at least in some classes) can be arranged to commute with the Clifford level. This makes the theory solvable level by level. We demonstrate this for the specific case of the uncolored $O(n)^3$ tensor model with arbitrary even $n$, and reduce the question of determining the spectrum and eigenstates to an algebraic equation relating Young tableaux. Solving this reduced problem is conceptually trivial and amounts to matching the representations on either side, as we demonstrate explicitly at low levels. At high levels, representations become bigger, but should still be tractable. None of our arguments require any supersymmetry.

\vspace{1.6 cm}
\vfill

\end{titlepage}

\setcounter{footnote}{0}


\section{Introduction}

Holography has provided us with our most complete definition of quantum gravity so far. This means that we know perfectly well-defined quantum mechanical systems which contain classical gravity in some appropriate  limit. See \cite{Rajesh, Papadodimas} for some discussions on the general philosophy of holography. However, having a definition of quantum gravity is not the same as being able to extract the quantum physics of gravity from it. For that, we need to not just know what the theory is, we also need to usefully solve it at non-zero values of Planck's constant. This, has largely remained a challenge.

Typically in holographic theories, we have a paramter $N$ which can loosely be thought of as the inverse of a (dimensionless) Planck's constant. The usual situation is that we have reasonable control on the theory in the classical $N \rightarrow \infty$  limit, and perhaps also perturbatively to low orders \cite{c=1} in $1/N$. In some cases \cite{doublescaled} we even know the perturbative quantum corrections to all orders in $1/N$. This means that we know at least some models were holographic theories are under reasonable perturbative control. To the best of our knowledge however, there is no known example where a holographic theory is {\em fully} solvable {\em exactly} for arbitrary tunable values of $N$. This is not just a matter of being pedantic, there are strong reasons \cite{Maldaeternal, CKQFTBHReview, Barbon, Hartman, Fitzpatrick} to think that unitarization of black holes (for example), will require\footnote{Let us emphasize a few things here, however: (1) The double-scaled matrix models of \cite{doublescaled} are dual to topological strings and do not contain black holes, so this criticism is not directly relevant to them. (2) It is not necessary that we be able to solve the theory {\em completely} at finite $N$ before we can answer {\em any} non-pertubative question about quantum gravity. For example, localization \cite{localization-reviews} has been useul in exactly evaluating some partition functions and operators non-perturbatively in contexts where supersymmetry is available. In this and other cases, some quantities are computable exactly. But what we mean by full solvability in this paper is that (in principle) we know all the correlation functions in the theory, or equivalently, that we know the spectrum and the eigenstates. (3) To answer detailed dynamical questions, especially regarding local bulk physics (like in the case of black hole unitarity etc.), it seems that a far more detailed understanding of non-perturbative questions than what we currently have access to, will be necessary. (4) The theories we discuss in this paper, and claim are exactly solvable, are expected to be dual to asymptotically AdS$_2$ spaces which have substantially limited dynamics \cite{Almheiri-Polchinski} than full higher dimensional gravity. So these theories are likely not general enough to fully answer questions related to black hole unitarity. But they are certainly dynamically more rich than the (quasi-)topological cases discussed in \cite{doublescaled}. They also exhibit saturated chaos \cite{MS, Polchinski-Rosenhaus} as expected in black hole physics \cite{MSS}, which is not the case in the matrix model \cite{blackhole-nonformation-Malda-Strom}, see also \cite{Polchinski-failed-matrix}.} one to understand non-perturbative corrections in $1/N$. 

In this paper, we will argue that the holographic tensor models that have recently emerged \cite{Witten, Klebanov, Tanasa, Bala1, Pavan, Gurau1, Gurau2,  Prithvi, Loga} in the context of SYK models \cite{dump} are a potential candidate for such fully solvable theories. See \cite{tensordump} for other related works on tensor models. 

Exact diagonalization of the holographic tensor models might seem like a hopeless task at first. Indeed, in \cite{Bala1,Pavan} these theories were numerically diagonalized for small (and fixed) values of some appropriately chosen $N$. But it was also observed that as the $N$ was increased, the matrix sizes grew exponentially as $\sim \exp(N^\#)$ where $\#$ is $\ge 1$, and depends on the specific model. This makes exact diagonalization essentially impossible fairly quickly even with modern computers. Also, a numerical diagonalization is not as useful for many crucial purposes from a holographic point of view. In particular, constructing singlets in order to make contact with a ``closed string" bulk perspective is much more intuitive when one has at least some analytic understanding. Furthermore, even in the simplest cases considered in \cite{Bala1, Pavan} explicit diagonalization on a computer was too demanding for the eigenvectors to be found: only the eigenvalues were found in \cite{Bala1, Pavan}.

Nontheless, in this paper we will argue that determining the eigenvalues and eigenvectors of Hamiltonians of (at least some classes of) holographic tensor models can be reduced to highly tractable combinatorical questions regarding representation decompositions. We will support this claim by simplifying the diagonalization problem for the even rank uncolored $O(n)^3$ tensor models using a basic observation: that the Hilbert space of these theories forms a spinor representation, and that by appropriate choice of the spinor raising and lowering operators, one can ensure that the Hamiltonian commutes with the Clifford level. The advantage of this is that it enables us to diagoinalize the Hamiltonian level by level, which in turn reduces it to a problem of breaking down tensor products of representations in the language of Young tableaux. We believe that a complete solution of this latter problem is possible, but it becomes cumbersome and requires a systematic approach when it comes to the representations at high levels, so we will present it in a forthcoming publication \cite{Bala2}. In this paper, we will derive the {\em statement} of the reduced problem in terms of Young tableaux (equation (\ref{maineq}) is our main result), as well as present its solution for Clifford levels 0, 1 and 2 (which automatically yields the solution for the highest, penultimate and pre-penultimate Clifford levels as well). The diagonalization problem reduces to the algebraic question of equating the representation content on either side of this equation. Solving  it is conceptually trivial, as our low level calculations will demonstrate. But at high levels will require working with big representations -- perhaps on a computer.

In the next two sections, we will present the set up and our results. In a final section, we make various comments on related questions, implications of our results and future directions.

\section{The Setup}

\subsection{Holographic Tensor Model}

The theory we consider in this paper is an uncolored tensor model, written down in \cite{Klebanov} adapting the colored model in \cite{Witten}. In particular, the Hamiltonian that we wish to diagonalize is:
\begin{align}
H&=\psi ^{ijk}\psi ^{ilm}\psi ^{pjm} \psi ^{plk}
\end{align} 
where each of the tensor indices take values from 1 to $n$. The theory is quantum mechanical, ie., there is only a time diection and no space. Note that contractions between different tensors can occur only between the same slots on each tensor. The components of the tensors are ``Majorana fernions", by which condensed matter theorists mean real objects that satisfy Clifford algebra, aka Gamma matrices. We have set the coupling to unity for conveneince, it can always be reinstated by dimensional analysis. This Hamiltonian, by construction, is $O(n)^3$ invariant. The fermionic tensors obey the following commutation relations:
\begin{align}
\{\psi ^{i_1j_1k_1},\psi ^{i_2j_2k_2}\}&=\delta ^{i_1i_2}\delta ^{j_1j_2}\delta ^{k_1k_2} \label{clifford}
\end{align} 
In the rest of the paper, we work with the case of even $n$. We believe that an adaptation of the technology we develop here should also exist for the odd $n$ case. But in this paper our goal is to demonstrate a proof of principle, so we will be minimalistic.

\subsection{Making Clifford Work}

For even $n$, we can define the following quantities:
\begin{align}
\Gamma ^{ij{k^\pm}}\equiv \psi ^{ij{k^\pm}}&=\frac{1}{\sqrt{2}}\left(\psi ^{ijk}\pm i \psi ^{ij(k+1)}\right)
\end{align}
where $k^{\pm}$ take values from 1 to $\frac{n}{2}$. These redefined tensors has the following anti-commutation relations:
\begin{align}
\{\psi ^{ij{k^+}},\psi ^{lm{n^+}}\}=0; \hspace{5 mm}\{\psi ^{ij{k^-}},\psi ^{lm{n^-}}\}=0; \hspace{5 mm} \{\psi ^{ij{k^+}},\psi ^{lm{n^-}}\}=\delta ^{il}\delta ^{jm}\delta ^{k^+,n^-}
\end{align} 
Since the Clifford algebra (\ref{clifford}) is realized in the Hilbert space of the theory, it is natural to try and rewrite the Hamiltonian in a form that takes advantage of it. Note that the  algebra (\ref{clifford}) is {\em not} the Clifford algebra of $SO(n)$, but of $SO(n^3)$, so one might think that working with Clifford raising and lowering operators that transform in $SU(n^3/2)$ might be beneficial. We have not been able to make too much use of this, however. Instead simplifications happen when one works with raising/lowering operators that transform in $O(n)^2 \times SU(n/2)$. This is perhaps not surprising because we suspect this is the biggest symmetry group of the Clifford raising/lowering operators that can be contained within the symmetry group $O(n)^3$ of the theory. 

In particular, the Hamiltonian can be written in terms of these redefined tensors as follows:
\begin{align}
H=&\sum \psi ^{ij{k^+}}\psi ^{il{m^+}}\psi ^{nj{m^-}}\psi ^{nl{k^-}}+\psi ^{ij{k^+}}\psi ^{il{m^-}}\psi ^{nj{m^+}}\psi ^{nl{k^-}} \nonumber \\
&+~~\psi ^{ij{k^-}}\psi ^{il{m^+}}\psi ^{nj{m^-}}\psi ^{nl{k^+}}+\psi ^{ij{k^-}}\psi ^{il{m^-}}\psi ^{nj{m^+}}\psi ^{nl{k^+}}
\end{align}
Note that a generic definition of raising/lowering operators will completely scramble the Hamiltonian, so let us take a moment to admire the fact that this form is in fact quite simple. In fact, this can be further simplified using the anti-commutation relations as follows:
\begin{align}
H&=\frac{n^4}{4}+\sum 2 \left(\psi ^{ij{k^+}}\psi ^{il{m^+}}\psi ^{nj{m^-}}\psi ^{nl{k^-}}-\psi ^{ij{k^+}}\psi ^{nj{m^+}}\psi ^{il{m^-}}\psi ^{nl{k^-}}\right)
\end{align}

\subsection{Level 0 and Level 1}

A Clifford representation is constructed by starting with the lowest weight state $\left|~\right\rangle$ to be annihilated by all the annihilation operators $\psi ^{ij{k^-}}$ i.e.,
\begin{align}
\psi ^{ij{k^-}} \left|~\right\rangle &=0
\end{align}
Note also that this condition immediately implies that the lowest weight state is an eigenstate of the Hamiltonian with the eigenvalue $\frac{n^4}{4}$. Thus we have already identified an eigenstate and its eigenvalue.

The rest of the Hilbert space can be generated by acting with the creation operators on $\left|~\right\rangle$. So, a general state at level $r$ can be written as:
\begin{align}
\psi ^{i_1j_1k^+_1}\psi ^{i_2j_2k^+_2}\ldots \psi ^{i_rj_rk^+_r}\left|~\right\rangle
\end{align}
Note that there are $ {n^3/2}\choose{r}$ such states possible. So, the total number of states at all $n^3/2$ levels together is equal to $2^{n^3/2}$. 

Now that we have a basis of states, we proceed to identify the eigenstates of the Hamiltonian. Before doing that, it is useful to note the following identity:
\begin{align}
\label{identity}
\psi ^{lmn^-}&\psi ^{ijk^-}\psi ^{i_1j_1k^+_1}\psi ^{i_2j_2k^+_2}\ldots \psi ^{i_rj_rk^+_r}\left|~\right\rangle \nonumber \\
&=\left(\delta ^{ii_1}\delta ^{jj_1}\delta ^{kk_1}\delta ^{li_2}\delta ^{mj_2}\delta ^{nk_2}-\delta ^{ii_2}\delta ^{jj_2}\delta ^{kk_2}\delta ^{li_1}\delta ^{mj_1}\delta ^{nk_1}\right)\psi ^{i_3j_3k^+_3}\psi ^{i_4j_4k^+_4}\ldots \psi ^{i_rj_rk^+_r}\left|~\right\rangle \nonumber \\
&+\left(\delta ^{ii_3}\delta ^{jj_3}\delta ^{kk_3}\delta ^{li_1}\delta ^{mj_1}\delta ^{nk_1}-\delta ^{ii_1}\delta ^{jj_1}\delta ^{kk_1}\delta ^{li_3}\delta ^{mj_3}\delta ^{nk_3}\right)\psi ^{i_2j_2k^+_2}\psi ^{i_4j_4k^+_4}\ldots \psi ^{i_rj_rk^+_r}\left|~\right\rangle \nonumber \\\
&+\left(\delta ^{ii_2}\delta ^{jj_2}\delta ^{kk_2}\delta ^{li_3}\delta ^{mj_3}\delta ^{nk_3}-\delta ^{ii_3}\delta ^{jj_3}\delta ^{kk_3}\delta ^{li_2}\delta ^{mj_2}\delta ^{nk_2}\right)\psi ^{i_1j_1k^+_1}\psi ^{i_4j_4k^+_4}\ldots \psi ^{i_rj_rk^+_r}\left|~\right\rangle \nonumber \\
&+\ldots
\end{align}
The equation above leads to a crucial fact:  namely that the Hamiltonian commutes with the Clifford level. It also immediately demonstrates that the generic level 1 states are also annihilated by the Hamiltonian, which means {\em all} of them are eigenstates with eigenvalue zero as well. 

In what follows, our basic strategy will be to use this fact to try and diagonalize the Hamiltonian level by level. 

\subsection{Level 2 and Higher}

To get some intuition, we will first identify the eigenstates at level $r=2$. Acting with the Hamiltonian on the state $\psi ^{i_1j_1k^+_1}\psi ^{i_2j_2k^+_2}\left|~\right\rangle$, we obtain the following:
\begin{align}
\label{H on r=2}
\left(H-\frac{n^4}{4}\right)\left(\psi ^{i_1j_1k^+_1}\psi ^{i_2j_2k^+_2}\left|~\right\rangle \right)&=4\left[\sum _i\psi ^{ij_2k^+_1}\psi ^{ij_1k^+_2}\delta ^{i_1i_2}-\sum _j\psi ^{i_2jk^+_1}\psi ^{i_1jk^+_2}\delta ^{j_1j_2}\right]\left|~\right\rangle
\end{align}
The above equation implies that the level 2 states with neither $i_1=i_2$ nor $j_1=j_2$ are eigenstates of the Hamiltonian with the same eigenvalue as that of the lowest weight state. This is the beginnings of an observation that we will apply systematically. The basic point is that the Hamiltonian is a singlet, and therefore if one can compare the representation content (under the $O(n)^2 \times SU(n/2)$ symmetry that is manifest) on both sides of the equation, we would have solved the problem of diagonalization.

Level 2 captures the essential structure of the action of the Hamiltonian on a level $r$ state. To see this we present the analogous equation at Clifford levels 3, for comparison.
The action of Hamiltonian on a generic state at level $r=3$ is given by:
\begin{align}
\label{H on r=3}
\left(H-\frac{n^4}{4}\right)&\left(\psi ^{i_1j_1k^+_1}\psi ^{i_2j_2k^+_2}\psi ^{i_3j_3k^+_3} \right)\left|~\right\rangle \nonumber \\
&= 4 \left[\sum _i \psi ^{ij_2k^+_1}\psi ^{ij_1k^+_2}\delta ^{i_1i_2}-\sum _j\psi ^{i_2jk^+_1}\psi ^{i_1jk^+_2}\delta ^{j_1j_2}\right]\psi ^{i_3j_3k^+_3}\left|~\right\rangle \nonumber \\
&+4\left[\sum _i \psi ^{ij_1k^+_3}\psi ^{ij_3k^+_1}\delta ^{i_3i_1}-\sum _j\psi ^{i_1jk^+_3}\psi ^{i_3jk^+_1}\delta ^{j_3j_1}\right]\psi ^{i_2j_2k^+_2}\left|~\right\rangle \nonumber \\
&+4\left[\sum _i \psi ^{ij_3k^+_2}\psi ^{ij_2k^+_3}\delta ^{i_2i_3}-\sum _j\psi ^{i_3jk^+_2}\psi ^{i_2jk^+_3}\delta ^{j_2j_3}\right]\psi ^{i_1j_1k^+_1}\left|~\right\rangle 
\end{align}
This can easily be generalized to general level. The point about these equations is that they relate representation content on either side. (The Hamiltonian is obviously a singlet under the symmetry of the theory.) So we will write the general equation in a representation theoretic language that makes this point immediate, and show that it is trivial to solve it if we know the representation breakdown. For level 2 this is easy, so we will do it explicitly. For higher and higher levels, the problem becomes more complicated.

\section{Diagonalization via Young Tableaux}

For a general level $m$, the action of the Hamiltonian is given by:
\begin{align}
\left(H-\frac{n^4}{4}\right)&\left(~\begin{ytableau}
i_1 &\none[\otimes] &i_2 &\none[\otimes]&\none[\dots]&\none[\otimes] &i_m 
\end{ytableau}~ \right.,
\left.~\begin{ytableau}
j_1 &\none[\otimes] &j_2 &\none[\otimes]&\none[\dots]&\none[\otimes] &j_m 
\end{ytableau}~,~ k_1,k_2,\ldots k_m \right) \nonumber \\
&=4\sum _{p<q} (-1)^{p+q-1} ~(n)~ \left[\left(\bullet _{i_pi_q}~,~ \begin{ytableau}j_q& \none[\otimes]&j_p \end{ytableau}~,~k_p,k_q\right)-\left(\begin{ytableau}i_q&\none[\otimes]&i_p \end{ytableau}~,~ \bullet _{j_pj_q}~,~ k_p,k_q\right)\right]  \nonumber \\
&\otimes\left(	\underbrace{~\begin{ytableau}
i_1 &\none[\otimes] &i_2 &\none[\otimes]&\none[\dots]&\none[\otimes] &i_m 
\end{ytableau}~}_{\text{no} ~i_p ~\& ~i_q}~,~ 
\underbrace{~\begin{ytableau}
j_1 &\none[\otimes] &j_2 &\none[\otimes]&\none[\dots]&\none[\otimes] &j_m 
\end{ytableau}~ }_{\text{no} ~j_p ~\& ~j_q}~,~ \underbrace{k_1,k_2,\ldots k_m}_{\text{no} ~k_p ~\& ~k_q}\right) \label{maineq}
\end{align} 
We have written the equation now by reverting to the representation theoretic language of Young tableaux. It is straightforward to check that this is correct by directly acting with the Hamiltonian at general level. It is also easy to check that it reduces to the previous expressions at level 2 and 3 (after translating to the previous notation). 

To explain the notation, and to demonstrate the general idea, we solve the level 2 case in this language. The indices are written according to $O(n)_{i}\times O(n)_{j}\times SU(n/2)_{k}$, where $i, j$ and $k$ refer to the first, second and third indices of the fermionic tensors.  Let us start with the equation \eqref{H on r=2} which we present here for convenience:
\begin{align}
\left(H-\frac{n^4}{4}\right)\left(\psi ^{i_1j_1k^+_1}\psi ^{i_2j_2k^+_2}\left|~\right\rangle \right)&=4\left[\sum _i\psi ^{ij_2k^+_1}\psi ^{ij_1k^+_2}\delta ^{i_1i_2}-\sum _j\psi ^{i_2jk^+_1}\psi ^{i_1jk^+_2}\delta ^{j_1j_2}\right]\left|~\right\rangle
\end{align}
The Hamiltonian is a scalar under $O(n)_{i}\times O(n)_{j}\times SU(n/2)_{k}$. So, the LHS in terms of Young tableaux is given by:

\ytableausetup{centertableaux}
\begin{align}
\left(\begin{ytableau}
i_1  &\none[\otimes]& i_2  
\end{ytableau}\right. ,
\left.\begin{ytableau}
j_1  &\none[\otimes]& j_2  
\end{ytableau}~, ~k_1,k_2 \right)
=\left(\begin{ytableau}
i_1 & i_2  &\none [+] 
\end{ytableau}
\begin{ytableau}
i_1  \\ i_2  
\end{ytableau}
+\bullet _{i_1i_2}
\right. ,
\left.\begin{ytableau}
j_1 & j_2  &\none [+] 
\end{ytableau}
\begin{ytableau}
j_1  \\ j_2  
\end{ytableau}
+ \bullet _{j_1j_2}~,~k_1,k_2
\right) 
\end{align}
where we have defined
\begin{align}
\bullet _{i_1i_2}&=\frac{\delta ^{i_1i_2}}{n}~\text{Tr}(~)^{ii} \nonumber
\end{align}
As far as the RHS is concerned, only some of the above diagrams show up and are given as:
\begin{align}
\left(n ~\bullet _{i_1i_2} ~,~\begin{ytableau}
j_2 & j_1  &\none [+] 
\end{ytableau}
\begin{ytableau}
j_2  \\ j_1  
\end{ytableau}
+ n~\bullet _{j_1j_2} ~,~ k_1,k_2
\right) -
\left(\begin{ytableau}
i_2 & i_1  &\none [+] 
\end{ytableau}
\begin{ytableau}
i_2  \\ i_1  
\end{ytableau}
+n ~\bullet _{i_1i_2}
 ~,~ n ~\bullet _{j_1j_2} ~,~k_1,k_2\right)
\end{align} 
Upon comparing the LHS and RHS, we can conclude that the following states have an eigenvalue of ``$+4n$'':
\begin{align}
\left(H-\frac{n^4}{4}\right) \left(\bullet _{i_1i_2}~,~\begin{ytableau}
j_1 & j_2  
\end{ytableau}~,~ k_1,k_2 \right)&=+4n \left(\bullet _{i_1i_2}~,~\begin{ytableau}
j_1 & j_2  
\end{ytableau}~,~ k_1,k_2 \right)\nonumber \\
\left(H-\frac{n^4}{4}\right) \left(~\begin{ytableau}
i_1 \\ i_2  
\end{ytableau}~,~\bullet _{j_1j_2}~,~k_1,k_2\right)&=+4n \left(~\begin{ytableau}
i_1 \\ i_2  
\end{ytableau}~,~\bullet _{j_1j_2}~,~k_1,k_2\right)
\end{align}
Further, the following two sets of states have an eigenvalue of ``$-4n$'':
\begin{align}
\left(H-\frac{n^4}{4}\right) \left(\bullet _{i_1i_2}~,~\begin{ytableau}
j_1 \\ j_2  
\end{ytableau}~,~ k_1,k_2\right)&=-4n \left(\bullet _{i_1i_2} ~,~ \begin{ytableau}
j_1 \\ j_2 
\end{ytableau}~,~ k_1,k_2\right)\nonumber \\
\left(H-\frac{n^4}{4}\right) \left(~\begin{ytableau}
i_1 & i_2  
\end{ytableau}~,~\bullet _{j_1j_2}~,~ k_1,k_2\right)&=-4n \left(~\begin{ytableau}
i_1 & i_2  
\end{ytableau}~,~\bullet _{j_1j_2}~,~ k_1,k_2\right)
\end{align}
The rest of the states have a zero eigenvalue. These results can also be re-done directly with more indices etc., and we show that in Appendix A for convenience in understanding our notation.

The idea then is to compare the representation content on either side of (\ref{maineq}) and that will yield the eigenstates directly.  Note that the eigenvalues and the eigenfunctions are all purely determined by representation theory at this stage. 

\subsection{Comments}

We note a few relevant points here, and leave a detailed study of the higher levels for future work.

What we have seen is that the diagonalization of the Hamiltonian reduces to a group theory problem, where we are instructed to match representations in (\ref{maineq}) level by level. The basic ingredients for combining representations in the Young tableaux language are the Littlewood-Richardson rules. Here we need them for the $O(n)^2 \times SU(n/2)$ group. The $SU(n/2)$ representation structure  of equation (\ref{maineq}) is trivial (modulo one caveat we will discuss), so the question is really about combining $O(n)\times O(n)$ representations. The Littlewood-Richardson rules in the $O(n)$ case are known\footnote{More precisely, $SO(n)$ case. But this difference only affects the details of the calculations, not the actual result.} \cite{Skvortsov} and so this is in principle straightforward. What we have done in level 2 is an example of such an operation. The trouble of course is that the representations grow as we go to higher and higher levels (even though they will turn around by level $n^3/4$ because of the finiteness of the Clifford representation).

This is straightforward enough, but there are two things one needs to be careful about. One is that the tensors we are combining here are fermions, so they should be treated as anti-commuting objects. This becomes important when the tensors that are being combined are to be treated as identical, which happens when the value of the $k_i$ indices on two of the tensors that are being combined happen to be the same. When that happens, some of the representations that result are identically zero. We illustrate this point for level 2, by explicitly showing the count of the number of states that fall into each representation in an appendix. 

The details of the ``plethysms" that we will need will be discussed in the forthcoming paper \cite{Bala2}.

\section{Conclusion}

Our explicit example here has been for the specific case of uncolored tensor models with a symmetry group of even rank. But the crucial ingredient in our strategy was more general: we exploited the interplay between the the level structure of the spinor representation and the Hamiltonian. We found that the two commute, which simplified the problem enormously. But this may not be necessary for solvability -- if the algebra of the Hamiltonian and the level have a suitable structure, there could be a systematic way to exploit it. It will be interesting to see if other holographic tensor models (in particular, the odd rank version of the theory considered here) lead to solvability in some such approach. Also, it seems highly plausible that a strategy like the one presented here might go through for the colored tensor models as well \cite{Bala2}.

Another comment worth making is that we have {\em not} used the full symmetries of the Hamiltonian. In particular, the full $O(n)^3$ symmetry of the Hamiltonian is not manifest in our Clifford representation approach. This raises two points.
\begin{itemize}
\item One should be able to check that the full spectrum that one obtains at the end has the full symmetry. (Note that the level construction is what breaks manifest symmetry, so one will not see the full symmetry realized in any one level alone.)
\item Is there a way to set up the problem so that one can exploit the symmetries more systematically than we have? 
\end{itemize}


One sinister way in which this program can fail (at least for the case of the even $n$ tensor models we consider here) is if the theory for even $n$ and odd $n$ are qualitatively different, and only the odd $n$ theory behaves like SYK in the large $n$ limit. The explicit case diagonalized in \cite{Pavan} and found to have some chaos-like behavior was an odd $n$ case, the higher even $n$ cases were inaccessible to numerical diaogonalization.  We find the scenario that the odd and even cases are qualitatively different to be quite unlikely (note also that in the colored model, the even case already exhibited chaos \cite{Bala1}), afterall it seems reasonable to expect that the large-$N$ limit of the even and odd cases should match. But we are not sure if this is a watertight way to rule out this malicious possibility, which is why we have decided to menion it here.

There is another well-known context \cite{Ramgoolam} in holography, where Young tableaux (and very closely related Schur polynomials) have emerged as a fundamental ingredient. This is in the reconstruction of BPS states/geometries in the bulk, via the gauge singlets in the boundary. Note that the Young tableaux that emerge in our construction are different. Roughly speaking, the construction of \cite{Ramgoolam} present the ``closed string" perspetive on the problem, while ours is an ``open string" picture. Of course, once one follows through the program we have outlined here to completion, it will be very interesting to identify the gauge singlet states and operators and to explicitly reconstruct the bulk. It seems highly plausible that such a construction will clarify the relationship between the two distinct avatars of Young tableaux in these theories.

We conclude with one further comment. In holographic theories,  one is only interested in gauge singlets on the boundary. One way to implement this is to note that after diagonalization, one only keeps the gauge singlet states. But one can also reverse the order of this process and look only for eigenstates that are gauge singlet states to start with. From that perspective, the problem we have tackled here is harder than it needs to be. We will discuss this issue further in \cite{Bala2}.

There are clearly numerous questions that one can address with a full solution of a holograohic quantum theory. We will exercise restraint in taking stock of such a wishlist here.

\section*{Acknowledgments}

CK thanks Justin Raj David and Ramalingam Loganayagam for discussions. We thank P. N. Bala Subramanian for a closely related collaboration \cite{Bala2}.

\appendix

\section{Explicit Diagonalization of Level 2}

In the main body of the paper, we presented the eigenstates and eignevalues at level 2 working directly with Young tableaux. In this appendix, we reinterpret the same calculation in terms of explicit creation/annihilation operators etc. at level 2.

A general state at level 2 is of the form $\psi ^{i_1j_1k_1}\psi ^{i_2j_2k_2}$. Hamiltonian acting on such a state gives us:
\begin{align}
\left(H-\frac{n^4}{4}\right)\left(\psi ^{i_1j_1k^+_1}\psi ^{i_2j_2k^+_2} \right)\left|~\right\rangle&=4\left[\sum _i\psi ^{ij_2k^+_1}\psi ^{ij_1k^+_2}\delta ^{i_1i_2}-\sum _j\psi ^{i_2jk^+_1}\psi ^{i_1jk^+_2}\delta ^{j_1j_2}\right]\left|~\right\rangle
\end{align}
We now take a trace over the indices $i_1$ and $i_2$ i.e., we multiply the above equation by $\delta ^{i_1i_2}$ and then take a sum over $i_1$ and $i_2$. Doing so, we will get:
\begin{align}
\left(H-\frac{n^4}{4}\right)\sum _i\left(\psi ^{ij_1k^+_1}\psi ^{ij_2k^+_2} \right)\left|~\right\rangle&=\left[4n~\sum _i\psi ^{ij_2k^+_1}\psi ^{ij_1k^+_2}-4\sum _{j,i}\psi ^{ijk^+_1}\psi ^{ijk^+_2}\delta ^{j_1j_2}\right]\left|~\right\rangle
\end{align}
Similarly, a trace over the indices $j_1$ and $j_2$ gives us:
\begin{align}
\left(H-\frac{n^4}{4}\right)\sum _j\left(\psi ^{i_1jk^+_1}\psi ^{i_2jk^+_2} \right)\left|~\right\rangle&=\left[4\sum _{i,j}\psi ^{ijk^+_1}\psi ^{ijk^+_2}\delta ^{i_1i_2}-4n~\sum _{j}\psi ^{i_2jk^+_1}\psi ^{i_1jk^+_2}\right]\left|~\right\rangle
\end{align}
Taking a linear combination of the above three equations, we get:
\begin{align}
\left(H-\frac{n^4}{4}\right) \left[\psi ^{i_1j_1k^+_1}\psi ^{i_2j_2k^+_2}-\frac{1}{n}\left(\sum _i\psi ^{ij_1k^+_1}\psi ^{ij_2k^+_2}\delta ^{i_1i_2} +\sum _j\psi ^{i_1jk^+_1}\psi ^{i_2jk^+_2}\delta ^{j_1j_2} \right) \right]\left|~\right\rangle&=0
\end{align} 
So, the traceless part of a generic second level state is an eigenstate of the Hamiltonian with an eigenvalue $\frac{n^4}{4}$. 	

Further, it can also be shown that
\begin{align}
\left(H-\frac{n^4}{4}\right) \sum _i\left(\psi ^{ij_1k^+_1}\psi ^{ij_2k^+_2}- \psi ^{ij_2k^+_1}\psi ^{ij_1k^+_2}\right)\left|~\right\rangle &= -4n \sum _i\left(\psi ^{ij_1k^+_1}\psi ^{ij_2k^+_2}- \psi ^{ij_2k^+_1}\psi ^{ij_1k^+_2}\right)\left|~\right\rangle \nonumber \\
\left(H-\frac{n^4}{4}\right) \sum _j\left(\psi ^{i_1jk^+_1}\psi ^{i_2jk^+_2}- \psi ^{i_2jk^+_1}\psi ^{i_1jk^+_2}\right)\left|~\right\rangle &= +4n \sum _j\left(\psi ^{i_1jk^+_1}\psi ^{i_2jk^+_2}- \psi ^{i_2jk^+_1}\psi ^{i_1jk^+_2}\right)\left|~\right\rangle
\end{align}
Taking a trace over both the $i$'s and $j$'s, we get an eigenstate with an eigenvalue $\frac{n^4}{4}$:
\begin{align}
\left(H-\frac{n^4}{4}\right) \sum _{i,j}\psi ^{ijk^+_1}\psi ^{ijk^+_2}\left|~\right\rangle &=0
\end{align}

\section{Counting at level 2}

In this appendix, we explicitly count the number of states that fall into various irreducible representations at level 2.  A generic state at this level is given by:
\begin{align}
\left(\begin{ytableau}
i_1  &\none[\otimes]& i_2  
\end{ytableau}\right. ,
\left.\begin{ytableau}
j_1  &\none[\otimes]& j_2  
\end{ytableau}~, ~k_1,k_2 \right)
\end{align}
 Note that the total number of states at this level are $n^3/2 \choose {2}$ as there are $\frac{n^3}{2}$ creation operators.

We consider the case of $k_1\neq k_2$. In this case all the possible irreps show up. So, the possible states are:
\begin{align}
\underbrace{\left(\begin{ytableau}
i_1 & i_2  &\none [+] 
\end{ytableau}
\begin{ytableau}
i_1  \\ i_2  
\end{ytableau}
+\bullet _{i_1i_2}
\right. ,
\left.\begin{ytableau}
j_1 & j_2  &\none [+] 
\end{ytableau}
\begin{ytableau}
j_1  \\ j_2  
\end{ytableau}
+ \bullet _{j_1j_2}~,~k_1,k_2
\right) }_{{n^2\times n^2 \times }{{n/2}\choose {2}}}
\end{align}
 Since all the irreps corresponding to $i$'s and $j$'s are present and $k_1\neq k_2$, we find that the total number of states with $k_1\neq k_2$ are $\frac{n^5}{8}(n-2)$.

The case of $k_1=k_2$ is somewhat tricky. Thanks to the antisymmetric nature of the $\psi $'s, some of the representations are trivially zero. The non trivial states with $k_1=k_2$ are  given by:
\begin{align}
\underbrace{\left(\begin{ytableau}
i_1 & i_2  &\none [+] 
\end{ytableau}n~\bullet _{i_1i_2}~,~ 
\begin{ytableau}
j_1  \\ j_2  
\end{ytableau}
 ~,~ k_1=k_2\right)}_{\frac{n(n+1)}{2}\frac{n(n-1)}{2}\frac{n}{2}}+
 \underbrace{\left(\begin{ytableau}
i_1 \\ i_2   
\end{ytableau}~,~ 
\begin{ytableau}
j_1  & j_2  &\none [+]
\end{ytableau}n~\bullet _{i_1i_2}
 ~,~ k_1=k_2\right)}_{\frac{n(n+1)}{2}\frac{n(n-1)}{2}\frac{n}{2}}
\end{align}
Since the symmetric traceless representation has the dimension of $\frac{n(n+1)}{2}-1$ and the antisymmetric part has the dimension of $\frac{n(n-1)}{2}$, we can see that the states with $k_1=k_2$ are $\frac{n^3}{4}(n^2-1)$ in number. Adding the contributions from both the cases gives the desired value of $n^3/2 \choose {2}$.

\end{document}